\documentclass[
  aps,prl,
  10pt,
  superscriptaddress,
  twocolumn, floatfix]{revtex4-1}

\usepackage[utf8]{inputenc}
\usepackage[T1]{fontenc}
\usepackage{lmodern}
\usepackage{xcolor}
\definecolor{linkblue}{RGB}{31,119,180}
\usepackage[
  unicode,
  colorlinks,
  citecolor=blue,
  linkcolor=linkblue,
  urlcolor=linkblue,
  bookmarks=true,
  bookmarksopen=true,
  bookmarksopenlevel=3,
  bookmarksnumbered=true]{hyperref}
\usepackage{siunitx}
\usepackage{graphicx}
\usepackage{amsmath}
\usepackage{amssymb}
\usepackage{fontawesome}

\newcommand{\maybe}[1]{\textcolor{black}{#1}}

\newcommand{\fig}[1]{Fig.~\ref{fig:#1}}

\newcommand{\USB}{\ensuremath{\omega_I}} 
\newcommand{\LSB}{\ensuremath{\omega_S}} 
\newcommand{\sqzfield}{\ensuremath{\omega_{S,I}}}
\newcommand{\LUSB}{\ensuremath{\omega_{LO,I}}} 
\newcommand{\LLSB}{\ensuremath{\omega_{LO,S}}} 
\newcommand{\pumpfield}{\ensuremath{\omega_{\rm pump}}} 
\newcommand{\auxfield}{\ensuremath{\omega_{\rm lock}}} 

\newcommand{\SqzFwhm}{\maybe{\SI{140}{MHz}}}
\newcommand{\FsrCavity}{\maybe{\SI{58.73}{MHz}}}
\newcommand{\LineWidthCavity}{\maybe{\SI{150}{kHz}}}
\newcommand{\firstdetuning}{\maybe{\SI{460}{kHz}}}
\newcommand{\symmetricdetuning}{\maybe{\SI{460}{kHz}}}

\begin{document}


\title{Demonstration of interferometer enhancement through EPR entanglement}

\author{Jan S\"udbeck}
\author{Sebastian Steinlechner}
\author{Mikhail Korobko}
\author{Roman Schnabel}
\affiliation{Institut f\"ur Laserphysik und Zentrum f\"ur Optische Quantentechnologien der Universit\"at Hamburg,\\%
Luruper Chaussee 149, 22761 Hamburg, Germany}

\maketitle

\textbf{
The sensitivity of laser interferometers used for the detection of gravitational waves (GWs) is limited by quantum noise of light.
An improvement is given by light with squeezed quantum uncertainties, as employed in the GW detector GEO\,600 since 2010.
To achieve simultaneous noise reduction at all signal frequencies, however, the spectrum of squeezed states needs to be processed by 100\,m-scale low-loss optical filter cavities in vacuum. Here, we report on the proof-of-principle of an interferometer setup that achieves the required processed squeezed spectrum by employing Einstein-Podolsky-Rosen (EPR) entangled states. Applied to GW detectors, the cost-intensive cavities would become obsolete, while the price to pay is a 3\,dB quantum penalty.
}

\textbf{Introduction} ---
The recent series of gravitational-wave (GW) detections by the Advanced LIGO and Advanced Virgo $2^{\rm nd}$ generation observatories \cite{GW150914,GWTC-1} launched the new field of GW astronomy. A $3^{\rm rd}$ generation (3G) of ground-based GW observatories aims at a signal-to-noise improvement by a factor of ten, resulting in an average increase in detection rate by three orders of magnitude and more precise localisations of the source. At these high sensitivities, binary neutron star mergers and their optical counterparts, acting as \emph{standard sirens} in the measurement of cosmological parameters \cite{HubbleConstant2017}, will be found on a daily basis. In addition, the ring-down phase of such mergers will reveal the so far unknown equation of state of neutron stars \cite{Abbott2017a}. A European design study for a 3G observatory, the \emph{Einstein Telescope} was already published in 2011 \cite{et_design}, since followed by proposals for similar efforts in the US \cite{ligo_white_paper_2018,abbott_exploring_2017}.

\begin{figure}[tb]
  \begin{center}
  \includegraphics[width=\linewidth]{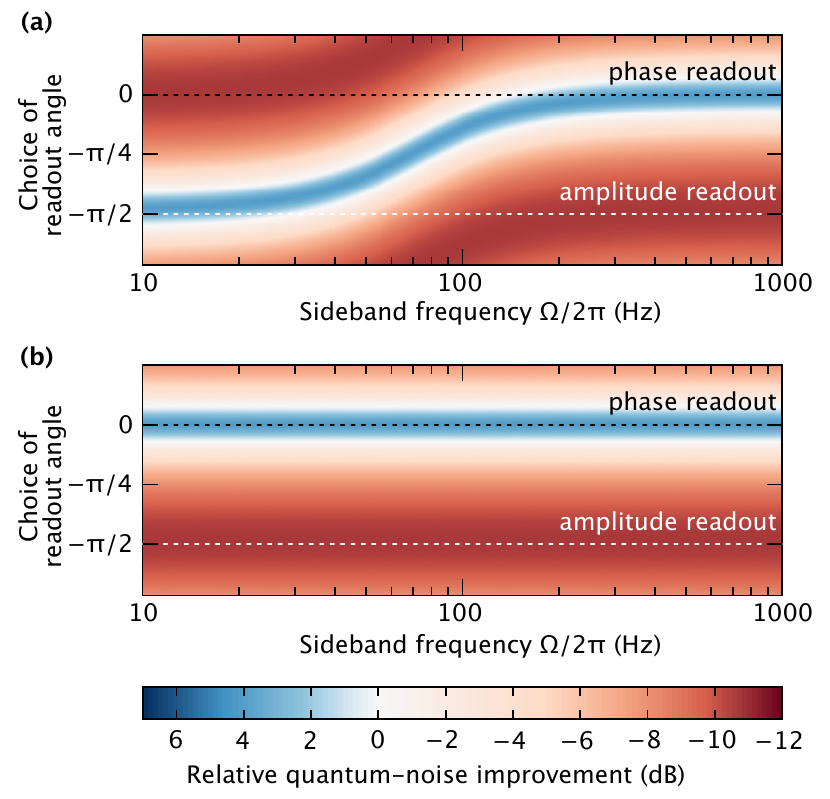}
  \end{center}
  \caption{Simulation of the improvement in readout quantum noise of a typical laser interferometer with squeezed light, as a function of signal frequency and readout angle, and with parameters similar to Advanced LIGO, including a loss factor \SI{40}{\%}. In the case of injected frequency-independent squeezing \textbf{(a)}, the noise in the signal (phase) readout is reduced at high sideband frequencies. However, quantum back-action (coupled via radiation pressure) leads to a rotation of the squeezed readout angle, worsening the readout at low frequencies (curved blue band). Broadband improvement \textbf{(b)} could be achieved with a frequency-dependent squeeze angle, which would compensate this rotation (horizontal blue band). So far, this necessitated a complicated hardware processing with additional filter cavities. The EPR scheme demonstrated here can generate such a frequency-independent spectrum without the need for these filter cavities.}
  \label{fig:qnoise}
\end{figure}%

To increase the measured GW signal strength, future ground-based observatories will have longer arms and will use higher light powers. Heavier mirrors, larger laser spots on the mirrors, cryo-cooled mirrors and possibly an underground location all help to reduce the zoo of different noise sources. Light with a squeezed uncertainty of its electric field \cite{Walls1983,Breitenbach1997,Schnabel2017} targets the reduction of interferometer quantum noise, separate from the light power \cite{Caves1981,Schnabel2010,McClelland2011}. Squeezing in one of the field quadratures, independently of the signal frequency, has been employed to reduce the observatory shot-noise \cite{LSC2011,LSC2013} and has been improving the sensitivity of GEO\,600 since 2010 \cite{Grote2013}.
Since the third observing run O3, it also is regularly used in the Advanced LIGO detectors and in Advanced Virgo. For a broadband reduction of both photon shot-noise (quantum measurement noise) as well as photon radiation pressure noise (quantum back-action noise), the squeezed quadrature needs to change from amplitude to phase quadrature depending on the signal frequency \cite{Jaekel1990,Danilishin2012}, cf.\ \fig{qnoise}. The optimal, \emph{frequency-dependent} squeeze spectrum can be produced by reflecting frequency-independent squeezed light off low-loss narrow-band \emph{filter cavities} \cite{Kimble2001,Chelkowski2005,Khalili2010}. Even with cavity mirrors of highest quality and operation in ultra-high vacuum, appropriate filter cavities must have lengths of the order of hundred meters to achieve the linewidth and loss requirements \cite{Barsotti2018}. Recently, Y.\,Ma and co-workers proposed to avoid these cost-intensive filter cavities \cite{Ma2017} by exploiting (i) the Einstein-Podolsky-Rosen (EPR) entanglement \cite{Einstein1935,Reid1989,Bowen2004} between different frequency components of the injected squeezed field \cite{Schori2002,Hage2010} and (ii) the signal-recycling cavity that is already part of current and future GW observatories. The same approach is also useful in detuned signal-recycled interferometers without radiation-pressure noise \cite{Brown2017}.

Here, we report on the experimental proof of principle of the proposal made in \cite{Ma2017}. In our experiment, a linear optical cavity represented the GW observatory signal-recycling cavity. Onto this cavity, we mode-matched a conventional, frequency-independent squeezed field. Employing bichromatic (two-frequency) balanced homodyne readout, we demonstrated that the unwanted effect on quantum noise caused by a detuning of the cavity can be undone by conditioning the measurement at the signal frequency on the measurement at the corresponding, EPR-correlated idler frequency. Our scheme corresponds to a realisation of the EPR gedanken experiment \cite{Einstein1935}, where a measurement at location $A$ instantaneously results in reality of the entangled quantity at location $B$.

\begin{figure}
  \center
  \includegraphics[scale=1.1]{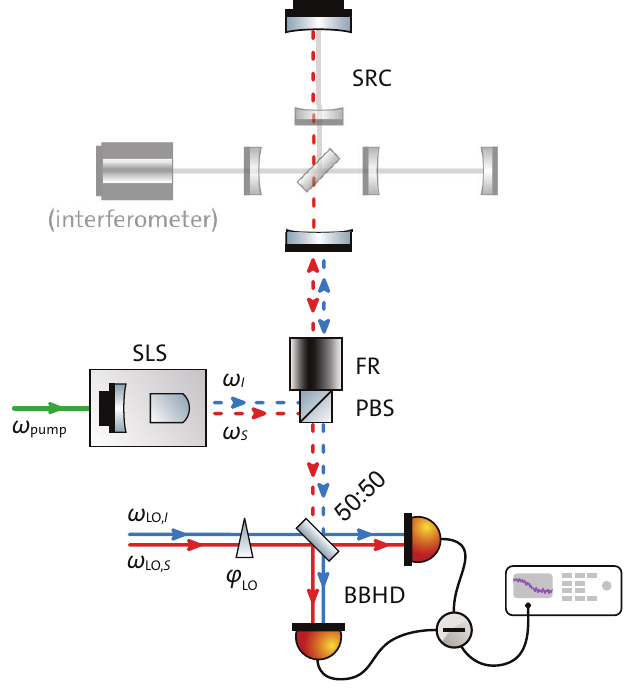}
  \caption{Simplified representation of this work's experimental setup. The frequency components emitted from a broadband squeezed-light source (SLS) that are reflected off a cavity experience a frequency dependent phase that can compensate for the frequency dependence of quantum back action. If the SLS is operated at particular centre frequency, the signal-recycling cavity (SRC) can be used that already exists in every GW observatory. The EPR entangled sidebands \USB and \LSB of the squeezed field are detected with a bichromatic balanced-homodyne detection (BBHD), employing local oscillators at frequencies \LLSB{} and \LUSB{}. The greyed-out parts indicate how the cavity of our experiment could relate to a full-scale measurement device (here, an Advanced LIGO-style Michelson interferometer with arm resonators, power-, and signal recycling). PBS, polarizing beam-splitter; FR, Faraday rotator. SRC length in our experiment was \SI{2.5}{m}.}
\label{fig:setup}
\end{figure}%

\begin{figure*}[ht]
  \center
  \includegraphics[width=\textwidth]{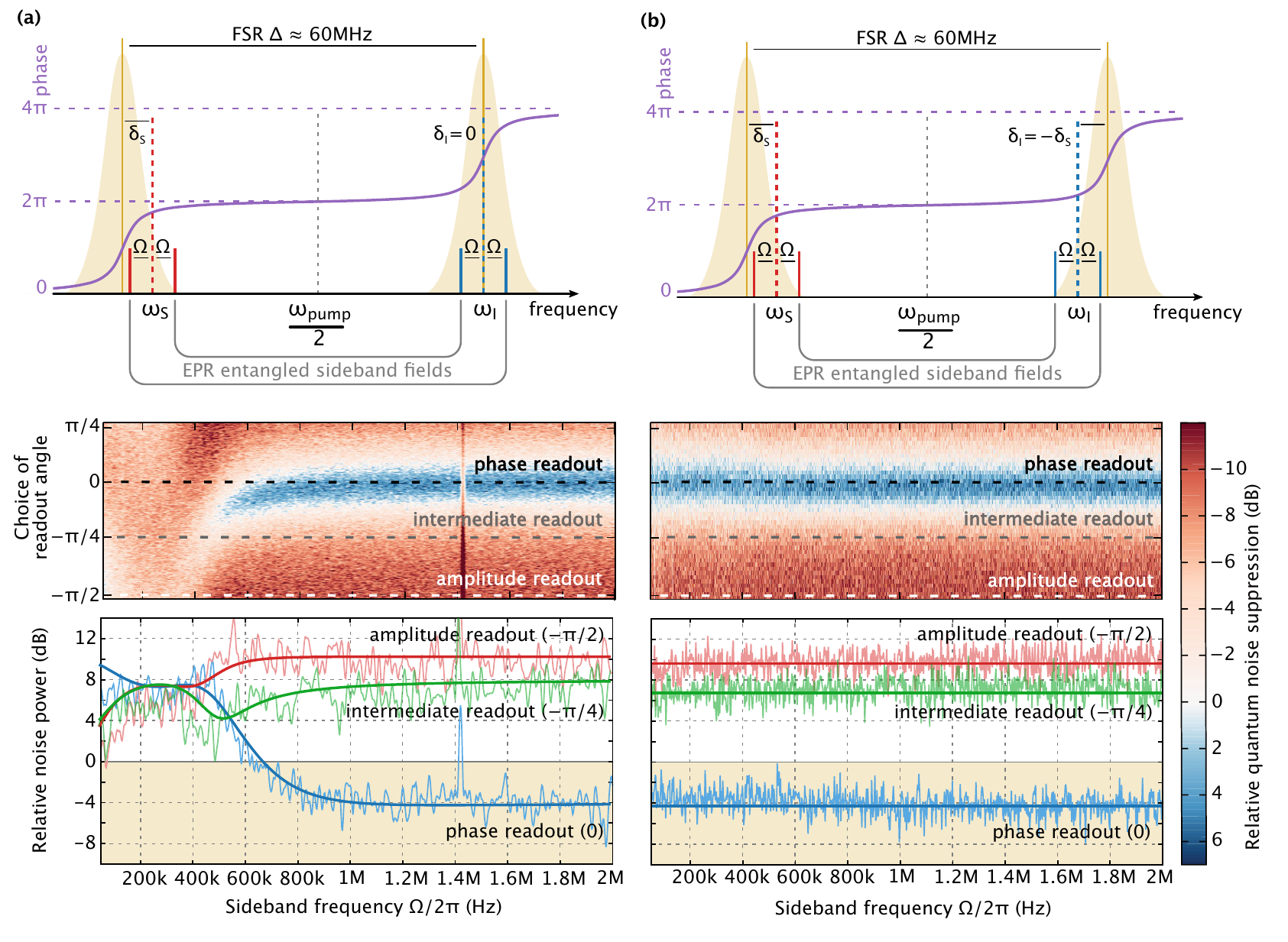}
  \caption{The schematics in the top row show the relative positions of signal and idler bands around \USB\,and \LSB\,in our demonstration of frequency-dependent squeeze angle rotation. Electric fields at frequencies symmetrically located around half the SLS pump frequency, $\pumpfield{}/2$, are mutually EPR entangled. In the bichromatic homodyne readout, electric fields at $\LSB\pm\Omega$ and $\USB\pm\Omega$ are combined in such a way that the result directly gives the total quantum noise of the detection scheme. The spectrograms in the middle show the measured quantum noise variances for continuously scanned readout angle. The bottom plots present cuts of the spectrogram at phase, intermediate and amplitude readout quadratures: light traces show experimental data; dark traces show theoretical fits with a joint set of parameters. \textbf{(a)} While the signal band around \LSB{} is detuned from cavity resonance by $\delta_S = 2\pi\times\firstdetuning$, the idler band around \USB{} is held exactly on the next resonance. For frequencies $\Omega$ much larger than the cavity linewidth and detuning, a noise level \SI{4}{dB} below shot noise is achieved in a specific readout quadrature. Around the detuned cavity resonances, however, sidebands of \LSB{} and \USB{} experience different phase shifts, spoiling the EPR correlations and resulting in an increased noise. The spectrogram shows a frequency-dependent squeezing phase at frequencies below $\sim1$\,MHz. The exact frequency dependence is a function of the detuning related to the cavity linewidth.  \textbf{(b)} When \LSB{} and \USB{} are detuned from cavity resonances by the same but opposite amount $\delta_S$, EPR entangled fields at all sidebands $\Omega$ receive the same frequency-dependent phase shift. As a result, the detrimental effect of detuning (and/or ponderomotive squeezing) on the detection quantum noise can be cancelled and a flat, frequency-independent squeezed spectrum is obtained. This validates the proposed scheme for quantum-noise enhancement with EPR-entangled states. The features at around \SI{50}{kHz}, \SI{100}{kHz} and \SI{1.2}{MHz} are experimental artefacts caused by our phase control scheme.}
\label{fig:signal_detuned}
\end{figure*}%

\begin{figure}[tb]
  \center
  \includegraphics[width=\linewidth]{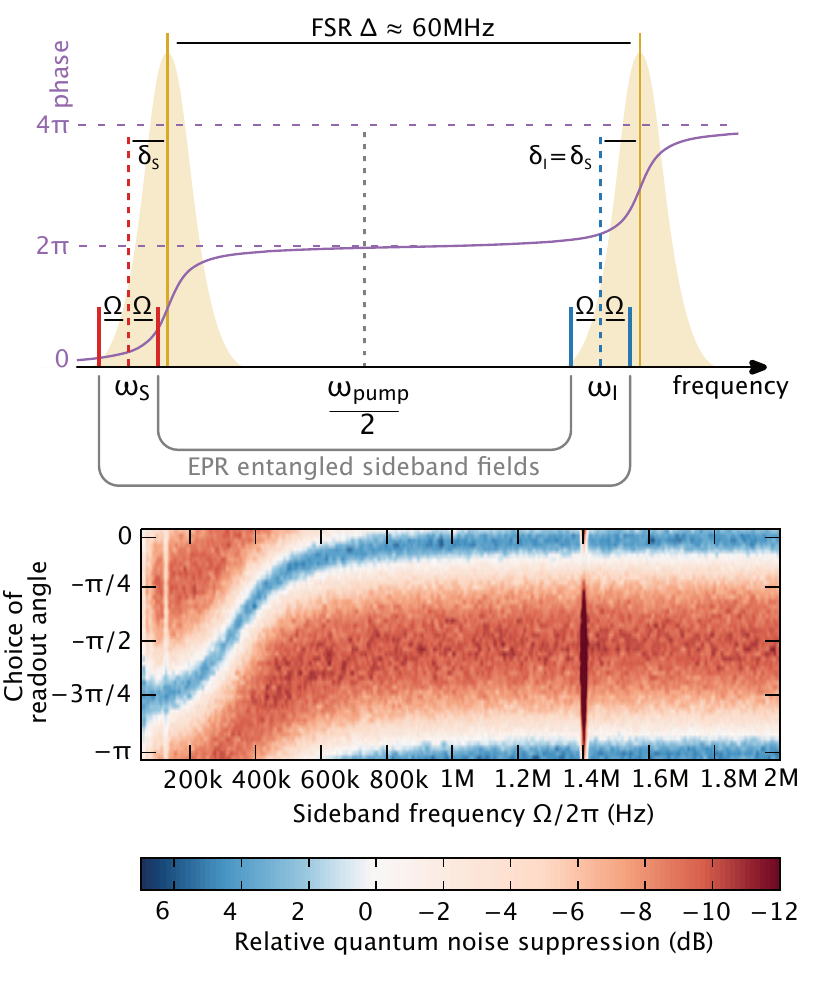}
  \caption{
    This figure shows the results of an additional experimental step demonstrating the flexibility of the setup. In this case both \LSB{} and \USB{} are detuned from cavity resonance in the \emph{same} direction and by the same amount $\delta_S=\delta_I=2\pi\times\symmetricdetuning$, leading to a frequency-dependent squeeze angle. The spectrogram shows the measured quantum noise variances for continuously scanned readout angle. The acquired frequency dependence in squeeze angle sweeps across the phase range of about $3\pi/4$. This is larger than needed in gravitational-wave detector for cancelling the ponderomotive squeezing effect, but shows the potential of the setup in tailoring the frequency dependence of squeeze angle to the needs of quantum-optical experiment. The exact frequency dependence is ultimately a function of the cavity linewidth and the detunings. The features at around \SI{50}{kHz}, \SI{100}{kHz} and \SI{1.2}{MHz} are experimental artefacts caused by our phase control scheme.
  }
  \label{fig:waterfall}
\end{figure}

\textbf{Quantum measurement process} ---
In interferometric gravitational-wave detectors, optomechanical interaction between the light and the test-mass mirrors induces correlations in quantum noise due to the ponderomotive squeezing effect \cite{Danilishin2012}: quantum fluctuations in the amplitude quadrature of the light create fluctuating radiation pressure forces on the suspended mirrors.
The resulting motion in turn couples into the phase quadrature of the reflected light, and the amplitude and phase quadratures of the light become quantum correlated.
Mathematically, this coupling defines the relation between the input and output quadratures in a  tuned interferometer (omitting the unimportant phase factor)\cite{Kimble2001}:
 \begin{align}
     &\hat{b}_{\rm out}^{(\rm ampl)} = \hat{b}_{\rm in}^{(\rm ampl)},\\
    &\hat{b}_{\rm out}^{(\rm phase)} = \hat{b}_{\rm in}^{(\rm phase)} - \mathcal{K}(\Omega)\,\hat{b}_{\rm in}^{(\rm ampl)} + \rm{signal}\label{eq:sig},
 \end{align}
 where the optomechanical coupling is described by the coupling strength (Kimble factor) $\mathcal{K}(\Omega)$ that depends on the signal frequency $\Omega$:
 \begin{equation}
     \mathcal{K}(\Omega) = \frac{2 J \gamma}{\Omega^2 (\gamma^2 + \Omega^2)}, \quad J = \frac{8 \pi I_c}{M \lambda L}.
 \end{equation}
The Kimble factor depends on the light power inside the arm cavities of the detector $I_c$ and their length $L$, as well as the on the optical linewidth of the detector $\gamma$, the mass of the mirrors $M$ and the laser wavelength $\lambda$.
Its frequency dependence has contributions from the cavity Lorentzian profile $\gamma^2 + \Omega^2$ and from the mechanical response of the free test mass $\Omega^2$.
The Kimble factor becomes larger at lower sideband frequencies, which is exactly the effect of radiation pressure noise contaminating the low-frequency sensitivity $\hat{b}_{\rm out}^{(\rm phase)} \approx - \mathcal{K}(\Omega)\hat{b}_{\rm in}^{(\rm ampl)} + \rm{signal}$.
The radiation pressure noise at low frequencies can therefore be reduced by squeezing the amplitude quadrature $\hat{b}_{\rm in}^{(\rm ampl)}$.
At the same time the shot noise in phase quadrature dominates the high frequency sensitivity: $\hat{b}_{\rm out}^{(\rm phase)} \approx \hat{b}_{\rm in}^{(\rm phase)}+ \rm{signal}$, and thus calls for squeezing the phase quadrature $\hat{b}_{\rm in}^{(\rm phase)}$.
This frequency dependence of the quantum noise in the interferometer is shown on the \fig{qnoise}(a), which illustrates the need for frequency-dependent squeezing.
When such frequency-dependent squeezing is injected from the outside, the detected noise in the phase quadrature takes the form
\begin{multline}\label{eq:phase_quadrature}
        \hat{b}_{\rm out}^{(\rm phase)} =
        \hat{b}_{\rm in}^{(\rm phase)} \bigl[\cosh r - \sinh r\left(\cos 2\phi + \mathcal{K} \sin 2\phi\right)\bigr] - \\ - \mathcal{K}\hat{b}_{\rm in}^{(\rm ampl)} \bigl[\cosh r + \sinh r\left(\cos 2\phi - \mathcal{K}^{-1}\sin 2\phi\right)\bigr],
\end{multline}

with squeeze factor $r$ and a frequency-dependent squeeze angle $\phi\equiv\phi(\Omega)$.
A straightforward optimization leads to $\phi(\Omega) = \arctan \mathcal{K}(\Omega)$, which results in optimally reduced noise at all frequencies, $\hat{b}_{\rm out}^{(\rm phase)} = e^{-r} \left(\hat{b}_{\rm in}^{(\rm phase)}-\mathcal{K}(\Omega)\hat{b}_{\rm in}^{(\rm ampl)}\right)$, compare with Eq.\eqref{eq:sig}.
The effect of the frequency-dependent squeezing on the quantum noise can be seen in \fig{qnoise}(b).

The frequency-dependent rotation of quadratures by radiation-pressure noise has the same effect on the quantum noise as detuning a cavity without movable mirrors from its resonance by $\delta$.
In this case the coupling constant in Eq.\,\eqref{eq:phase_quadrature} is replaced by $\mathcal{K}^{\rm cav}(\Omega, \delta) = 2\gamma \delta / \left(\gamma^2 - \delta^2 + \Omega^2 \right)$~\cite{Danilishin2012}.
While generally the frequency dependence of the two coupling factors is different, in the case of optimal detuning $\delta = \gamma$ within the cavity linewidth $\gamma \gg \Omega$ they become equivalent up to a scaling factor: $\mathcal{K}(\Omega) \approx 2 J /(\gamma \Omega^2) = \mathcal{K}^{\rm cav}(\Omega, \gamma) J/\gamma^3$.
This symmetry between a detuned cavity and the ponderomotive squeezing effect explains the need for a detuned filter cavity for producing the frequency-dependent squeezing that reduces the quantum noise at all frequencies.
When squeezing is reflected off a detuned filter cavity, the frequency dependence that its angle acquires can be exactly opposite to the one caused by the radiation pressure inside the detector, and the two cancel each other.
This results in the frequency-independent sensitivity enhancement, as illustrated in \fig{qnoise}(b).
Such a frequency dependence is possible to create with EPR-entangled states of light, following the proposal by Ma et al. \cite{Ma2017}.
While we do not demonstrate the ponderomotive squeezing effect in our experiment, the above relations allow us to emulate it with a detuned cavity.


\textbf{Generation of EPR entanglement} --- We produced light with a squeezed quantum uncertainty by sub-threshold pumping of a non-degenerate cavity-enhanced parametric amplifier. It naturally consisted of mutually entangled electric fields at lower (\emph{signal} $\LSB$) and higher (\emph{idler} $\USB$) sidebands of the central frequency, which was half the pump frequency $\pumpfield/2 = \USB{} + \LSB{}$.
Measuring the quadrature noise at $\sqzfield\pm\Omega$ individually with two separate homodyne detectors leads to an increased noise over the vacuum uncertainty. However, the measurement outcome at one detector could be \emph{conditioned} (i.e.\ in post-processing) on the measurement outcome of the other detector with an optional scaling factor $g$.
Then, for sufficiently high squeeze factors, the conditional variances $\hat{a}_s^{\rm (phase)} + g\,\hat{a}_i^{\rm (phase)}$ and $\hat{a}_s^{\rm (ampl)} + g\,\hat{a}_i^{\rm (ampl)}$ could be reduced to below the vacuum uncertainty (\emph{conditional squeezing}), certifying Einstein-Podolsky-Rosen entanglement \cite{Reid1989}.


\textbf{Experimental realization} ---
For our experimental demonstration of interferometer enhancement through EPR entanglement, we employed a setup as shown in \fig{setup}. A squeezed-light source provided a broadband squeeze spectrum with a full-width half-maximum of \SqzFwhm{}. Inside this linewidth, fields at $\pumpfield{}/2\pm\Delta\omega$ show mutually strong EPR correlations, of which we select the signal and idler bands around \LSB{} and \USB{} for our discussion. A \SI{2.5}{m}-long standing-wave cavity with free spectral range of $\Delta = \FsrCavity{}$ mimicked the signal recycling cavity of a GW detector. The cavity could be locked to arbitrary detunings with an auxiliary field \auxfield{} that was orthogonally polarized such as to not contaminate the measurement output. The squeezed field \sqzfield{} was mode-matched onto the linear cavity and the back-reflected light was separated with a Faraday rotator and a polarizing beam-splitter. We then performed bichromatic balanced homodyne detection \cite{Marino2007,Li2017} by overlapping the output field with two bright fields \LUSB{} and \LLSB{} at a 50:50 beam splitter and detecting the difference in photo current of the two outputs. For technical details on the setup and its implementation, see Methods.

In the first step, see \fig{signal_detuned}(a), we adjusted the EPR entangled field \sqzfield{} such that \LSB{} was slightly detuned by an offset frequency $\delta_S = \firstdetuning$ from a resonance of the linear cavity, while \USB{} was exactly on the next longitudinal resonance peak, one FSR $\Delta$ away. In this configuration, measurement sidebands $\pm\Omega$ around \LSB{} received a different (unequal and opposite) phase shift, leading to a frequency-dependent quadrature rotation via the coupling factor $\mathcal{K}^{\rm cav}(\Omega, \delta_S)$. On the other hand, the sidebands around \USB{} received equal and opposite phase shifts, resulting in the absence of quadrature rotation, $\mathcal{K}^{\rm cav}(\Omega, \delta_I=0) = 0$. At the balanced-homodyne detector, quantum noise components at $\LSB{}\pm\Omega$ and $\USB\pm\Omega$ were detected with frequency-independent gain and for a full $2\pi$ sweep of the local oscillator angle $\phi_{LO}$. The resulting photo-current variance is shown in \fig{signal_detuned}(a), normalized to the vacuum noise of both local oscillators. In accordance with the EPR \emph{gedanken experiment}, a noise power below \SI{0}{dB} demonstrates that it was possible to infer the noise components around \LSB{} by conditioning on a measurement at \USB{}. Here, this was the case for sideband frequencies $\Omega$ that were much larger than the linewidth of the linear cavity. In effect, the original squeeze factor was restored. Around the cavity linewidth, however, the squeeze factor was lost at all readout angles, since the frequency rotation around \LSB{} meant that the quantum correlations at the signal and idler wavelengths were no longer aligned for optimal inference. As explained in Ma et al. \cite{Ma2017}, an optimal Wiener filter would still allow to recover (most of) the correlations, but it cannot be implemented in our bichromatic BHD readout scheme.

In the second step, \fig{signal_detuned}(b), we detuned \USB{} away from the cavity resonance as well, by $\delta_I = -\delta_S$. In this configuration, the noise sidebands around \USB{} experienced equal, but opposite phase shifts to the noise sidebands around \LSB{}. As a result, the two measurement bases were optimally aligned at all frequencies, thus cancelling the frequency-dependent rotation everywhere. The resulting spectrum shows a constant squeeze factor for a single readout quadrature, demonstrating near-perfect inference of the quantum noise components around \LSB{} from the components around \USB{}.

Finally, we performed an additional step, not motivated by the gravitational-wave detection directly, but demonstrating the flexibility and the potential of the experimental setup. For that we detuned \USB{} away from the cavity resonance in another direction, by $\delta_I = \delta_S$, see \fig{waterfall}. As a result, the noise sidebands around \USB{} were phase shifted by the same amount as the ones around \LSB{}. The resulting squeezing spectrum acquired frequency dependence, spanning a larger phase range than in the first step of the experiment. This demonstrates the possibility to create a desired frequency-dependent squeezing spectrum by carefully tuning the filter cavities.


\textbf{Conclusion} ---
Since the third observation run, the Advanced LIGO and Advanced Virgo observatories benefit from frequency-independently squeezed states of light to reduce the quantum shot noise at high signal frequencies.
Due to the quantum measurement process, both observatories will suffer from increased quantum back-action noise at low signal frequencies. Additional long-baseline, low-loss filter cavities are considered for resolving this issue. Here, we provide the proof of principle that the same effect can be achieved with the signal-recycling cavity, which exists already in GW observatories. Following the proposal in \cite{Ma2017}, we utilized Einstein-Podolsky-Rosen entanglement.
As pointed out in \cite{Ma2017}, the approach suffers from a \SI{3}{dB} noise penalty on the initial squeezing and an increased sensitivity to optical loss, both of which are due to the measurement of an additional pair of sidebands. However, in light of the high additional costs for low-loss, narrow-linewidth filter cavities, the demonstrated broadband enhancement of interferometer sensitivity through EPR entanglement can still be a viable alternative.


\begin{acknowledgments}
This work was supported by the Deutsche Forschungsgemeinschaft (DFG), project SCHN757/6-1. It has the LIGO document number P1900226.
\end{acknowledgments}



\textbf{Methods} ---
\emph{Squeezed-light source:} We used a monolithic cavity-enhanced OPA to generate broadband squeezed states of light via parametric down-conversion in PPKTP. The cavity was pumped by \SI{100}{mW} of light at $\omega_\mathrm{pump}=2\pi c/(\SI{1064}{nm}/2)$.

\emph{Linear cavity:} We set up a \SI{2.5}{\metre}-long linear cavity for emulating the signal-recycling cavity of a GW detector resulting in a free spectral range of \FsrCavity. The mirror on the back of the cavity had a highly-reflective coating and the coupling mirror had a reflectivity of \SI{97}{\%} resulting in a linewidth of $\sim$\LineWidthCavity{} (HWHM). The cavity was locked with the Pound-Drever-Hall method using an auxiliary field in the orthogonal polarization. We used the zero-crossing of one sideband of the resulting error signal to allow for an easy shift of the resonance frequency.

\emph{Bichromatic LO:} The bichromatic local oscillator was created by a strongly phase modulated light of the frequency $\omega_\mathrm{pump}/2$ with an electro-optic modulator. Afterwards, the upper and lower sidebands were each extracted from the carrier by filter cavities. Both sidebands were overlapped on a 50/50 beamsplitter and sent towards the detector. This method guaranteed a symmetric generation around the center frequency of the squeezed field and the detection at entangled sideband frequencies. We used a power of \SI{4}{mW} in the combined bichromatic oscillator with equal powers in both sidebands.

\emph{Theoretical modeling: }
We compute the input-output relations for the light fields propagating in our setup in a similar way to \cite{Ma2017, Brown2017}.
The resulting spectral density of the shot noise with frequency-dependent squeezing embedded is described by:
\begin{multline}
    S(\Omega) = 1 - \eta + \eta \cosh 2r + \\ + 2 \eta \frac{\sqrt{\alpha \beta} \sinh 2r}{\alpha + \beta} \left(\mathcal{K}_1(\Omega)\cos 2\zeta + \mathcal{K}_2(\Omega)\sin 2\zeta\right),
\end{multline}
where we defined the readout efficiency $\eta$, squeeze parameter $r$, readout angle $\zeta$, the power of local oscillators $\alpha, \beta$, and two coupling coefficients $\mathcal{K}_{1,2}(\Omega)$:
\begin{align}
    &\mathcal{K}_1(\Omega) =  \frac{\mathcal{C}(\Omega)}{\mathcal{D}(\Omega)}\left(\mathcal{C}(\Omega) - 2\gamma^2(\delta_1+\delta_2)^2\right),\\
    &\mathcal{K}_2(\Omega) =  \frac{\mathcal{C}(\Omega)}{\mathcal{D}(\Omega)}\left(2\gamma\left(\delta_1+\delta_2\right)\left(\gamma^2 - \delta_1\delta_2 + \Omega^2\right)\right),\\
    &\mathcal{C}(\Omega) = \left(\delta_1^2 - \Omega^2\right)\left(\delta_2^2 - \Omega^2\right)+\gamma^2\left(\gamma^2 +\delta_1^2 + \delta_2^2 + 2\Omega^2\right),\\
    &\mathcal{D}(\Omega) = \prod_{i=1,2}\left(\gamma^2 + \left(\delta_i-\Omega\right)^2\right)\left(\gamma^2 + \left(\delta_i+\Omega\right)^2\right).
\end{align}
This equation can be used to describe the detected spectra, as shown in Fig.\ref{fig:signal_detuned}.

\textbf{Author Contributions} ---
JS, SS and RS planned the experiment. JS and SS built and performed the experiment. MK provided the theoretical analysis. JS, SS, MK and RS prepared the manuscript.

\bibliography{EPR_ifo.bib}

\end{document}